\documentclass[aps,pre,twocolumn,superscriptaddress]{revtex4}
\usepackage[latin1]{inputenc}
\usepackage{textcomp}
\usepackage{amsmath}
\usepackage{amssymb}
\usepackage{graphicx}
\usepackage{wasysym}
\usepackage{physics}
\usepackage{stackrel}
\usepackage{multirow}
\usepackage{braket,mleftright}
\usepackage{empheq}
\usepackage{subfigure}
\usepackage{soul}
\usepackage{blkarray}
\usepackage{bm}
\usepackage{bbold}
\usepackage{color}
\usepackage[unicode=true,
 bookmarks=false,
 breaklinks=false,pdfborder={0 0 1},backref=false,colorlinks=false]
 {hyperref} 
\usepackage{breakurl}
\usepackage{dsfont}
\makeatletter

\usepackage{amsfonts}
\usepackage{amsthm}
\usepackage{graphics}
\usepackage{times}
\usepackage{bm}
\usepackage{color}
\usepackage{latexsym}
\usepackage{subfigure}
\usepackage{braket}
\usepackage{float}
\usepackage{palatino}
\usepackage{bbm}
\usepackage{color}

\newcommand{\beq}{\begin{equation}}
\newcommand{\eeq}{\end{equation}}
\newcommand{\bse}{\begin{subequations}}
\newcommand{\ese}{\end{subequations}}
\newcommand{\bea}{\begin{eqnarray}}
\newcommand{\eea}{\end{eqnarray}}

\makeatother

\begin{document}

\title{Work statistics for sudden quenches in interacting quantum many-body systems}

\author{Eric G. Arrais}
\email{eric.arrais@gmail.com}
\affiliation{Instituto de F\'{i}sica, Universidade Federal do Rio de Janeiro, 21941-972, Rio de Janeiro, Brazil}

\author{Diego A. Wisniacki}
\email{wisniacki@gmail.com}
\affiliation{Departamento de F\'{i}sica ``J. J. Giambiagi'' and IFIBA, FCEyN, Universidad de Buenos Aires, 1428 Buenos Aires, Argentina}

\author{Augusto J. Roncaglia}
\email{augusto@df.uba.ar}
\affiliation{Departamento de F\'{i}sica ``J. J. Giambiagi'' and IFIBA, FCEyN, Universidad de Buenos Aires, 1428 Buenos Aires, Argentina}

\author{Fabricio Toscano}
\email{toscano@if.ufrj.br}
\affiliation{Instituto de F\'{i}sica, Universidade Federal do Rio de Janeiro, 21941-972, Rio de Janeiro, Brazil}

\begin{abstract}

Work in isolated quantum systems
is a random variable and its probability distribution
function obeys the celebrated 
fluctuation theorems of Crooks and Jarzynski.
In this study, we provide a simple way to describe the work probability distribution function for sudden quench processes 
in quantum systems with large Hilbert spaces. 
This description can be constructed from two elements: the  level density of the initial Hamiltonian,
and a smoothed strength function 
that provides information about the influence of the perturbation
over the eigenvectors in the quench process, and is especially suited to describe quantum many-body interacting systems.
We also show how random models can be used to find such smoothed work probability distribution
and apply this approach to different one-dimensional spin-$1/2$ chain models. 
Our findings provide an accurate description of the work distribution of such systems in the cases of intermediate and high
temperatures in both chaotic and integrable regimes.

\end{abstract}
\maketitle

\section{Introduction}
\label{SectionI}

The past decade gave birth to a new research field, 
called quantum thermodynamics, that 
gathers concepts borrowed from two older fields, namely, thermodynamics and quantum mechanics. 
Much of the progress made on quantum thermodynamics concerns the interplay between quantum information and thermodynamics 
\cite{Goold2016,vinjanampathy2016quantum}, implementations of small-scale thermal machines 
\cite{Scovil1959, Alicki1979, Uzdin2015, Klaers2017, Peterson2018, deAssis2019}, 
the mechanisms behind the thermalization 
of quantum systems \cite{Srednicki1994, Linden2009, Gogolin2016, Borgonovi2016}, 
and the quantum fluctuation relations and nonequilibrium response of 
quantum many-body systems 
\cite{Silva2008, Dorner2012, Mascarenhas2014, Fusco2014, Wang2018, Zheng2018, Skelt2019, Zawadzki2019, Goold2018}. 
The latter is the main subject of the present study.
\par
Fluctuation theorems \cite{Jarzynski1997, Crooks1999,Esposito2009, HanggiReview} are simple equalities 
that establish relations between nonequilibrium quantities, such as the work, with equilibrium ones, such as the free energy.
These theorems were initially derived for classical systems and later extended, along with the definition of work, 
to the quantum regime.
While for initial diagonal states, such as thermal ones, there is consensus about the definition of quantum work in terms of the  
two projective energy measurement scheme 
\cite{Kurchan2000,Tasaki2000, Mukamel2003,Talkner2007,Roncaglia2014,Talkner2016}, there are some
desired properties that are not present when applying this definition to initial states with coherences in the energy 
eigenbasis \cite{PerarnauLlobet2017}.
Thus, there is still an ongoing debate about the proper definition of quantum work in the presence of coherences \cite{baumer2018fluctuating}.
Interestingly, the work statistics, and thus the verification of fluctuation theorems, 
were experimentally obtained both in the classical \cite{Liphardt2002, Collin2005, Blickle2006} 
and quantum domains \cite{Batalhao2014,An2015, Cerisola2017, Xiong2018, Rene2018}.

In the present study, we are interested in obtaining the work statistics for high-dimensional quantum many-body systems 
in thermal equilibrium after a sudden quench between the initial Hamiltonian
and a final one. We show that for large dimensional complex systems, for instance systems described by a random model, 
the work probability distribution function (pdf) admits a simple representation that does not require the full knowledge of the exact 
initial and final energy spectra and eigenvectors.
The work pdf can be obtained in terms of two smoothed energy functions: the first one is the level density,  
and the second one is a smooth version of the strength function (SF), also known as the local density of states
\cite{Flambaum2000}. We also show the conditions that random models representing the initial and final Hamiltonians must obey
in order to derive an ensemble average (EA) of the work distribution that has the same structure 
of the energy-smoothed work distribution found.
\par
We test our approach in two different models of one-dimensional spin-$1/2$ 
chain models. The first model is completely integrable while the second presents a transition from an integrable to a chaotic regime 
that can be described by a particular random model, called embedded Gaussian ortogonal ensembles EGOE$(1+2)$  \cite{Kota2001report},
where $(1+2)$ refers to Hamiltonians composed by terms with one and two-body interactions.
These systems are relevant in condensed matter physics and can be  
implemented in different experimental setups \cite{Duan2003, Chen2011, Simon2011}.
We show that the smoothed description is accurate in the regime of intermediate and high temperatures
for all perturbation strengths, irrespective to the system's integrability.
These findings provide a unified description of sudden quenches in interacting quantum many-body systems along with 
analytical expressions for the work pdfs. 

This paper is organized as follows. In Sec. \ref{SectionII} we briefly derive 
the work distribution associated to the definition of quantum work via the  two projective energy measurement scheme in sudden quenches. 
Then, in Sec. \ref{SectionIII}, we introduce the expression of the energy-smoothed work pdf for sudden quenches 
in systems with large Hilbert spaces.
In Sec. \ref{section-random-models} we derive an analogous expression for the EA work pdf of random Hamiltonian models.  
In Sec. \ref{TBRE} we discuss the level density and the strength function in the EGOE$(1+2)$, which are used in the description of the
deterministic spin-$1/2$ chain models considered in Sec. \ref{section-models}.
We apply our approach to these models in Sec. \ref{section-wpdf-models}.
Finally, in Sec. \ref{SectionVI} we present our conclusions.

\section{Work distribution for sudden quenches}
\label{SectionII}

We consider a work protocol where a system with Hamiltonian, $H:=H(0)$, is driven
to the final one, $\tilde H:=H(\tau)$, through the change of a controllable parameter described by a unitary $U_{\tau}$.
The initial state of the system is thermal at temperature $T$,
\begin{equation}
\varrho_{0} = \frac{e^{-\beta H}}{\mathcal{Z}_0}\,,
\label{eq:thermal}
\end{equation}
with $\mathcal{Z}_0 = \mbox{Tr}\,e^{-\beta H(0)}$ being the partition function for the initial Hamiltonian,  and 
$\beta :=1/k_B\,T$ with $k_B$ being the Boltzmann constant. In this context, 
work is defined as the difference in energy obtained from two  projective energy measurements,
one at the beginning and one at the end of the protocol.
Thus,  work is a random variable $w$ that can take values $w_{nm}:= \tilde{E}_{m} - E_{n}\,$, given by the possible energy differences,
where $\tilde{E}_{m}$ ($E_n$) is the energy of the $m$-th ($n$-th) level of the final (initial) Hamiltonian.
Defining $\Pi_{n}^{0}$ and $\Pi_{m}^{\tau}$ 
as the projectors over the $n$ and $m$ levels of the initial and final Hamiltonian eigenbases, respectively, 
we can compute the probability of obtaining $w_{nm}$ in a single run as 
$p_{m,n} = p_{n}p_{m|n}$. Here, $p_{n} = \mbox{Tr}\left[\varrho_{0}\Pi_{n}^{0}\right]$ is the probability of obtaining $E_{n}$ 
in the first projective measurement and
$p_{m|n} = \mbox{Tr}\left[\Pi_{m}^{\tau}U_{\tau}\Pi_{n}^{0}\varrho_{0}\Pi_{n}^{0}U_{\tau}^{\dagger}\right]/p_{n}$
is the conditional probability of obtaining $\tilde{E}_{m}$ in the second measurement given that the first measurement 
gave $E_{n}$ as a result.
Therefore, the pdf associated to the quantum work, called work distribution function, is
given by
\begin{equation}
P(w) := \sum_{m,n=1}^N p_{m,n}\, \delta\left[w - (\tilde{E}_{m} - E_{n})\right] \;.
\label{defPw}
\end{equation}
where $\delta$ represents the Dirac delta distribution, and $N$ is the dimension of the system.
For initial thermal states, such pdf obeys important fluctuation theorems, establishing   
relations between nonequilibrium and equilibrium properties of the system under consideration. One such theorem is 
the Jarzynski equality \cite{Jarzynski1997} that can be derived from
Crooks theorem \cite{Crooks1999}, which relates forward and backward protocols and shows that
both processes always differ by an exponential of the entropy production.
\par
Different kinds of driving protocols could be considered, but in this study we are interested in a 
specific one called sudden quench. This is represented by an instantaneous change of the Hamiltonian,
and thus the unitary operation associated with the driving is  represented by the identity operator, $U_{\tau} \approx \mathds{1}$.
We consider initial and final energy spectra with possible degeneracies, therefore
$\Pi_{n}^{0} := \sum_{\gamma}|\psi_n^{\gamma} \rangle \langle \psi_n^{\gamma}|$
and $\Pi_{m}^{\tau} := \sum_{\alpha}|\tilde{\psi}_m^{\alpha} \rangle \langle \tilde{\psi}_m^{\alpha}|$, where $\{|\psi_n^{\gamma} \rangle\}$ 
and $\{|\tilde{\psi}_m^{\alpha} \rangle \}$ are the sets of eigenvectors of $H$ and $\tilde{H}$, 
respectively, with eigenenergies $E_n$ and $\tilde E_m$, respectively, and $\gamma$ and $\alpha$, account for the possible degeneracies.
One can notice that we can rewrite the work distribution function as
\bea
P(w) &=& \sum_{n=1}^N \sum_{\gamma} \frac{e^{-\beta E_n}}{\mathcal{Z}_{0}} 
\;\mbox{SF}_{n,\gamma} (w)\,, \nonumber \\
\label{PWquench}
\eea
where ${\cal Z}_0 = \sum_{n=1}^N g_n \,e^{-\beta E_n}$
\footnote{The factor $g_n$ takes into account the possible degeneracies of the eigenenergy levels of the initial Hamiltonian $H$.}, 
and  $\mbox{SF}_n$ is the strength function of the nth eigenstate
of the initial Hamiltonian in terms of the final one, which is defined by
\bea
\mbox{SF}_{n,\gamma} (w) &:=& \sum_{m=1}^N \sum_{\alpha} |\langle \tilde \psi^{\alpha}_m | \psi^{\gamma}_n \rangle |^{2} \; \delta\left[w - (\tilde{E}_{m} - E_n) \right]. \nonumber \\
\label{strength}
\eea 
The $\mbox{SF}$ is the distribution of the squared modulus of the overlaps between 
initial eigenstates and final ones, and gives information about the effect produced by the quench on the system's eigenstates.
In the literature, it is also known as local spectral density or local density of states \cite{Casati1993}, and 
was introduced a long time ago in nuclear physics \cite{Wigner1955,Borhbook1969}. More 
recently, it was also used as a key concept in the definition of chaotic eigenstates of quantum many-body systems  
for the study of the eigenstate thermalization hypothesis \cite{Borgonovi2016}.
The $\mbox{SF}$ is normalized,
\bea
\int dw \;\mbox{SF}_{n,\gamma} (w) = \sum_{m}\sum_{\alpha}  \,| \ip*{\tilde{\psi}_m^{\alpha}}{\psi_n^{\gamma}}|^2 = 1 \,,
\eea
and its centroid is given by

\beq
\label{centespec}
\bar \varepsilon(E_n) = \int dw \;\mbox{SF}_{n,\gamma}\ (w)\;w=\tilde H^{\gamma\gamma}_{nn}  - E_n \,,
\eeq
where $\tilde H^{\gamma\gamma}_{nn}:=\langle \psi_n^\gamma | \tilde {H} | \psi_n^\gamma \rangle$ are the diagonal 
elements of the final Hamiltonian in the eigenbasis of the initial one. 
The variance of the $\mbox{SF}$ can be calculated by summing up the off-diagonal elements of the final Hamiltonian in 
the eigenbasis of the initial one,
\bea
\label{varespec}
\bar{\sigma}^2(E_n) &=& \langle \psi_n | \tilde {H}^2 | \psi_n \rangle - \langle \psi_n | \tilde {H} | \psi_n \rangle^2  \nonumber\\
&=&\sum_{n \neq n'} \sum_{\gamma\neq \gamma^\prime} |\tilde{H}_{nn'}^{\gamma\gamma^\prime}|^2 \,.
\eea
Remarkably, both quantities, centroid and variance, can be calculated without diagonalizing the final Hamiltonian,  
a task that could be hard in large many-body interacting systems.

\section{Work distribution for sudden quenches in systems with large Hilbert spaces}
\label{SectionIII}
\par
When one is dealing with complex systems with large Hilbert spaces, $N \gg 1$, like in quantum many-body systems, 
the spacing between nearest neighbor energy levels decreases with the dimension, so it is convenient to treat
the initial and final spectra as if they were continuous variables. Then, we can perform the following approximation
in Eq. \eqref{strength}:
\bea \label{infitinylimit} 
\sum_{m=1}^N\sum_{\alpha} (\ldots) \approx \int_{-\infty}^{\infty} \; d\tilde{E} \; \tilde{\rho}(\tilde{E}) \sum_{\alpha} (\ldots)\,,   
\eea
where $\tilde{\rho}(\tilde{E})$ is the level density of the final Hamiltonian, normalized to the total number of eigenlevels,
\bea \label{normdens}
\int_{-\infty}^{\infty} \; d\tilde{E} \; \tilde{\rho}(\tilde{E}) = N \;.
\eea
Using the approximation of Eq.~\eqref{infitinylimit}, 
we can write a smoothed version of the $\mbox{SF}$ of Eq.~\eqref{strength},
\bea \label{stfw}
\mbox{SF}_{\gamma}(w,E) &=& \int_{-\infty}^{\infty} d\tilde{E} \; \tilde{\rho}(\tilde{E}) \;\sum_{\alpha} |\ip*{\tilde{\psi^{\alpha}}(\tilde E)}{\psi^{\gamma}(E)}|^2  \nonumber\\
&&\times\delta\left[w - (\tilde{E} - E)\right] \nonumber \\
&=& \tilde{\rho}(w + E) \;\sum_{\alpha} |\ip*{\tilde{\psi^{\alpha}}(w+E)}{\psi^{\gamma}(E)}|^2 \,, \nonumber\\
\eea
where the notation $\ket{\psi^{\gamma}(E)}$ ($\ket{\tilde\psi^{\alpha}(\tilde E)}$) corresponds
to the energy eigenstate of $H$ ($\tilde H$) with energy $E$ ($\tilde E$). Note that 
the continuous version of the SF in Eq. \eqref{stfw} is correctly normalized. 
Using the approximation given by Eq. \eqref{infitinylimit} and the continuous version of the \mbox{SF} 
in Eq. \eqref{PWquench},  we obtain
\bea
P_{\rm sm}(w) &=& \frac{1}{\mathcal{Z}_{0}} \int^{\infty}_{-\infty} dE \; \rho(E) \;  e^{-\beta E} \;\sum_{\gamma} \mbox{SF}_{\gamma}(w,E) \,, \nonumber\\
\label{PWfinal}
\eea 
where the normalization is just the partition function associated with the initial thermal state
${\mathcal{Z}_{0}=\int^{\infty}_{-\infty} \,dE\,g(E)\, \rho(E)\,e^{-\beta E}}$.
\par
The approximation introduced in Eq. \eqref{infitinylimit} represents a sort of energy smoothing of the spectrum 
considered. In the limit of infinitely small energy smoothing, Eq. \eqref{PWfinal} recovers the work pdf in Eq. \eqref{PWquench}. 
This can be easily checked by using the level density
corresponding to an energy comb, $\rho(E)=\sum_{n = 1}^N\,\delta(E-E_n)$, in Eq. \eqref{PWfinal}, and an equivalent expression for
$\tilde\rho(\tilde E)$ in  Eq. \eqref{stfw}.
We shall show that the main behavior of the exact work pdf,
Eq. \eqref{PWquench}, of some  many-body systems can be described through 
an energy smoothing of the spectrum. This happens whenever the exact work pdf
fluctuates a little around Eq.~\eqref{PWfinal}, constructed from the
smoothed energy functions, $\rho(E)$ and $\mbox{SF}_{\gamma}(w,E)$. In such cases, the smoothed work pdf, $P_{\rm sm}(w)$, 
represents an advantage since, 
in principle, one does not need to know either the spectra or the eigenvectors of the Hamiltonians to describe 
$\rho(E)$  and $\mbox{SF}_{\gamma}(w,E)$. From the experimental point of view, this represents 
a viable alternative for systems with large Hilbert spaces, where the determination of the exact spectra 
could be impossible. From the theoretical point of view, the analytical determination of $P_{\rm sm}(w)$ involves 
the development of models that allow one to describe 
the smoothed functions, $\rho(E)$ and $\mbox{SF}_{\gamma}(w,E)$, in the systems with large Hilbert spaces,
without knowledge of the exact spectra of the initial and final Hamiltonians.
In many-body interacting systems, an important class of such models is the random models for the
Hamiltonian matrices of the quench. The next section is devoted to analyzing what are the features that
these random models must have in order to describe $P_{\rm sm}(w)$. 
\par


\section{Random models}

\par
In interacting many-body systems, the functional form of $\rho(E)$  and $\mbox{SF}_{\gamma}(w,E)$ 
could be difficult to obtain~\cite{Quirin2019}. 
A possible approach to deal with this problem is to verify whether the initial and final Hamiltonians can be fit into some 
random model. Therefore, we analyze the generic properties that random models representing the above class 
of Hamiltonians must have in order to provide an EA of the work pdf 
that matches the structure of the energy-smoothed work pdf in Eq. \eqref{PWfinal}. 
\subsection{The EA of the work distribution }
\label{section-random-models}
\par
In the study of the so-called quantum chaos and thermalization of isolated many-body 
systems, one often invokes some random model description of the Hamiltonians~\cite{Borgonovi2016}. 
They have also brought new insights to quantum thermodynamics
\cite{Halpern2018, Lobejko2017, Arrais2018}, with special attention devoted to the work pdf 
\cite{Lobejko2017, Arrais2018, Chenu2018}, and also were explored in the context of 
information scrambling \cite{Torres-Herrera2018, Chenu2019, Xu2018}.
Here, we show the basic conditions that any large Hilbert space random model has to fulfill in order to possibly obtain an expression of 
the work pdf with the structure of Eq. \eqref{PWfinal}.
\par
Any random model can be characterized by a joint probability distribution
 $P({\bf E},\boldsymbol{\theta})$ of energies and eigenvectors.
The random variables characterizing each model are the vector ${\bf E}:= (E_1,\ldots,E_N)$ 
$(\tilde {\bf E}:= (\tilde E_1,\ldots,\tilde E_N))$ containing the unperturbed (perturbed) eigenenergies, and
$\boldsymbol{\theta}$ $(\tilde{\boldsymbol{\theta}})$ 
that is the vector of parameters defining the initial (final) eigenvectors, 
viz. $\{\ket{\psi^{\gamma}_n(E_n,\boldsymbol{\theta})}\}$ ($\{\ket{\tilde{\psi}^{\alpha}_m(\tilde E_m,\tilde{\boldsymbol{\theta}})}\}$). 
We define
$P_{\rm EA}(w):=\langle\langle\langle\langle
P(w,{\bf E},\boldsymbol{\theta},\tilde{\bf E},\tilde{\boldsymbol{\theta}})
\rangle\rangle_{{\bf E},\boldsymbol{\theta}}\rangle \rangle_{\tilde{\bf E},\tilde{\boldsymbol{\theta}}}\,$ as the EA of 
the work pdf, $P(w,{\bf E},\boldsymbol{\theta},\tilde{\bf E},\tilde{\boldsymbol{\theta}})\,$, given by Eq. \eqref{PWquench}. We denote
the EA with respect to the initial ensemble of Hamiltonians by
\bea
\expval{\expval{(\ldots)}}_{{\bf E},{\boldsymbol{\theta}}}&:=&\int d{\bf E} \int d{\boldsymbol{\theta}}\,(\ldots)\, 
P({\bf E} ,\boldsymbol{\theta})\,,
\eea
and analogously for the final ensemble of Hamiltonians, $\expval{\expval{(\ldots)}}_{\tilde{\bf E},\tilde{\boldsymbol{\theta}}}\,$. 
All energy integrations are in the domain $(-\infty,\infty)$. Here, we do not assume any particular form of the ensembles of 
random Hamiltonians, but rather we want to show under which kind of assumptions
$P_{\rm EA}(w)$ has the structure of $P_{\rm sm}(w)$ in Eq. \eqref{PWfinal}.
\par
In Appendix \ref{appendixA} we show that the EA work pdf can be written as
\beq
\label{PEAw}
P_{\rm EA}(w) \approx  \frac{1}{\mathcal{Z}_{0, \rm EA}} \int^{\infty}_{-\infty} dE \; \rho_{\rm E}(E) \, 
e^{-\beta E} \; \sum_{\gamma} \mbox{SF}_{\gamma,\rm EA}(w,E)\,,
\eeq
under the following conditions: (i) statistical equivalence of the eigenvalues and 
eigenvectors of the Hamiltonians of the quench, (ii) an annealing approximation that is generically valid for 
large values of $N$  \cite{Cotler2017, Arrais2018, Xu2018, Chenu2019},  and (iii) the joint distributions approximately 
factorize as independent functions of the eigenlevels 
and eigenvectors, {\it viz.}
$P({\bf E}, {\boldsymbol{\theta}})\approx P({\bf E}) P( {\boldsymbol{\theta}})$.
To write Eq. \eqref{PEAw}, we defined the level density of the initial ensemble of Hamiltonians $H$, 
\beq
\rho_{\rm E}(E):=N\,\int d{\bf E}^\prime \, P({\bf E}),
\eeq
with $d{\bf E}^\prime$ meaning an integration over all eigenergies except one, viz. $E$.
We also defined the EA of the strength function as
\bea
\mbox{SF}_{\gamma,\rm EA}(w,E)&:=&\tilde \rho_{\rm E}(E+w) \nonumber\\
 &\times &
\sum_{\alpha} \langle\langle
 |\ip*{\tilde{\psi^{\alpha}}(w+E,\tilde{\boldsymbol{\theta})}}{\psi^{\gamma}(E,\boldsymbol{\theta})}|^2
 \rangle_{\boldsymbol{\theta}}\rangle_{\tilde{\boldsymbol{\theta}}}\,, \nonumber\\
 \label{SFEA}
\eea
with $\langle(\ldots)\rangle_{\boldsymbol{\theta}}:= \int d{\boldsymbol{\theta}}(\ldots) P({\boldsymbol{\theta}})$ (equivalently for 
$\langle(\ldots)\rangle_{\tilde{\boldsymbol{\theta}}}$) and the density of states of the final ensemble of Hamiltonians,
$\tilde \rho_{\rm E}(\tilde E = E + w)\,$. 
Finally, the EA partition function associated to the initial thermal state is
$\expval{{\cal Z}_0({\bf E})}_{\bf E}:=\mathcal{Z}_{0, \rm EA}:=\int^\infty_{-\infty} dE \,g(E)\,e^{-\beta E}\,\rho_{\rm E}(E)\,$.
Again, this partition function is just a normalization constant and allows one to check the 
consistency of the approximations made.
Therefore, using the  statistical independence of eigenvalues and eigenvectors
of the random ensembles of initial and final Hamiltonian matrices and the annealing approximation of the EA, 
we obtained an expression for the work pdf, Eq.~\eqref{PEAw}, that is equivalent to Eq.~\eqref{PWfinal}.
\par
The expression in Eq.~\eqref{PEAw} is useful only when the ensembles of Hamiltonians considered have
a sort of ergodic property, i.e., for a sufficiently large dimension $N$ the work pdf calculated from single draws of Hamiltonians from the 
ensembles (running average) is close to the EA  \cite{Weidenmuller2009}. 
This property was verified recently in \cite{Arrais2018} for the work pdf of quenches 
using the usual Gaussian ensembles (GEs)
of random matrix theory (RMT) \cite{Mehta-book, Weidenmuller2009}.
In Appendix \ref{Gaussian-ensembles} we rederive  
the results of \cite{Arrais2018} for Gaussian ensembles, but using Eq. \eqref{PEAw}. 
\par
\subsection{Two-body random ensembles}
\label{TBRE}
Many developments concerning the relation between the definition of quantum work and the classical definition
have already been accomplished \cite{Jarzynski2015, Wang2017, Pan2019}, with special attention devoted to quantum systems with 
classically chaotic counterparts \cite{Zhu2016, Garcia-Mata2017-1, Garcia-Mata2017-2}. 
Unlike in classical mechanics, chaos in quantum mechanics has no clear definition \cite{Stockmann-book}. However, 
analytical and numerical results suggest that quantum systems with classical chaotic counterparts share a common behavior for some 
of their spectral features \cite{Bohigas1984, Berry1985}.
For instance, the statistical behavior of the nearest-neighbor spacing distribution, $P(s)$ with $s:=E_{j+1}-E_j$, and the 
spectral rigidity, $\Delta_3$, follow the behavior predicted by the Gaussian ensembles of RMT
corresponding to quantum systems without time reversal invariance (GUE) 
or with time reversal invariance (GOE or GSE) \footnote{For the nearest-neighbor spacing distribution of unfolded spectra, $P(s)$ is 
well represented by the so-called Wigner's surmise \cite{Weidenmuller2009}.},
with the prominent characteristic of invariance with respect to basis rotations \cite{Mehta-book,Weidenmuller2009}.
For classically integrable quantum systems the corresponding behavior of these local spectral fluctuations is that of 
Poisson systems \cite{Berry1985}.
\par
Much more subtle is the concept of chaos in complex systems with no classical counterpart, 
which is common in the case of interacting quantum many-body systems \cite{Santos2012,Borgonovi2016}. 
For such systems, the chaotic regime 
is usually defined as the one where the level spacing distribution follows the prediction of the Gaussian ensembles.
However, it has been observed that the notion of quantum chaos is also directly related to the structure of the energy eingenstates.
This observation arose in the context of the study of
thermalization in isolated quantum systems of interacting particles, where the Hamiltonians can be separated into a sum of 
two parts, $H = H^\prime + \lambda\, V\,$. The first one, $H^\prime$, describes the non-interacting particles, or the ``mean field" part, 
being an effective single body Hamiltonian that could contain also some mean field effect coming from the two-body interactions, 
while the second one, $V$, describes the perturbation part, absorbing the typical two-body  interactions \cite{Santos2012,Borgonovi2016}. 
The important tool to characterize the structure of the energy eigenstates of the total Hamiltonian is 
the SF \cite{Wigner1955}. It gives the overlap distribution of a given eigenstate of $H$ with the mean field
basis of $H^\prime$. Typically, the eigenstates of $H$ may be considered chaoticlike 
for large values of the interaction strength, $\lambda$, where the SF displays a Gaussian form \cite{Santos2012}.
\par
Despite the success  of Gaussian ensembles in describing local fluctuations,  it became clear that no
realistic quantum system follows all the predictions of these ensembles.
In particular, the semicircle law for the level density, valid for Gaussian ensembles of large dimension, 
is completely artificial and is not followed by any physical system \cite{Stockmann-book}. 
Therefore, random models 
that consider the $k$-body nature of the interactions were introduced 
\cite{French1970, Bohigas1971, Bohigas1971-2, French1971, French1972, French1975}
in order to describe complex systems with realistic level densities. 
The most common $k$-body random ensembles are those with two-body interactions ($k=2$), called two-body random ensembles, 
which was recently used to address the increase of correlations in bosonic systems after a quench \cite{Borgonovi2019}, 
and in the analysis of spectral properties of a deterministic integrable quantum system \cite{Mailoud2019}. 
Within these ensembles, the most studied are the  embedded Gaussian ensembles (EGE), 
constructed by using an usual Gaussian ensemble in the $k$-particle space of effective interaction and 
then propagating it to the total $N$-particle space ($N>k$) using the direct product structure 
\cite{Kota2001report}. When one considers the GOE in the $k$-particle space, the associated embedded ensemble is known as EGOE.
\par
EGOEs exist for fermionic and also for bosonic systems \cite{Kotabook} and the EGOE$(1+2)$ are particularly useful for $(1+2)$-body 
Hamiltonians of the form
\beq
\label{H+V}
H = H^\prime + \lambda V\,,
\eeq
where $H^\prime$ is the one-body part, $V$ is the two-body residual interaction,
and $\lambda$ is the strength of the perturbation \cite{Kota2001report,Kotabook}.
$H^\prime$ describes the noninteracting constituents, particles, or quasi particles,
and in general comes from some mean-field approximation that defines the single-particle states filled with a number of 
non-interacting particles. Usually $H^\prime$ in 
EGOE$(1+2)$ are considered diagonal with or without random entries \cite{Borgonovi2016,Kota2001report,Kotabook}. The interactions 
between constituents are embedded into $V$, which contains the two-body nature of the interactions.
In the mean-field eigenbasis, the interaction term gives an approximate band like structure to the matrix $H$. 
Thus, the privileged character of the mean-field basis breaks the invariance of the EGOE$(1+2)$ 
with respect to basis rotations. Also, this basis is associated to the 
main property of quantum chaos in complex systems described by Hamiltonians 
of the form of Eq. \eqref{H+V}: the eigenstates of $H$ are delocalized in the mean-field basis
\cite{Borgonovi2016}.
\par 
One of the main features of the EGOE$(1+2)$ is that for a large dimension $N$, the level density associated to $H$ in Eq. \eqref{H+V}
displays a Gaussian form for every value of $\lambda$ \cite{Kota2001report,Kotabook}:
 \begin{equation}\label{dens1}
\rho_{\rm E}(E) \approx \rho_{\rm E}^{\rm G}(E) := \frac{N}{\sqrt{2\pi}\sigma_{\rm E}} e^{-(E -\bar{E})^{2}/2\sigma_{\rm E}^2}\,,  
\end{equation}
where the centroid $\bar{E}$, and the variance $\sigma^2_{\rm E}$ are independent of $N\,$.
Besides their dependence on the parameter $\lambda$, their values are also fixed  by the variance $\sigma^2$ of the 
Gaussian distributed diagonal entries of the Gaussian ensemble used in the $k$-particle space. The parameter $\sigma^2$  
defines the energy scale of the EGOE$(1+2)$ considered. Usually, the numerical values of the centroid and variance, 
$\bar{E}$ and $\sigma^2_E$, are obtained from a numerical fitting over the level density calculated from several single draws 
of Hamiltonians matrices of fixed large dimension $N$. The dependence of $\bar{E}$ and $\sigma^2_E$ on $N$ can be inferred by
repeating the procedure for various values of $N$ for which the diagonalization of the matrices is possible, and then extrapolating
for very large values of $N$.
\par
In the EGOE$(1+2)$, the strength function of the nth eigenstate of an unperturbed Hamiltonian, $H^\prime$, in the basis of a
perturbed one Eq.~\eqref{H+V}, depends on the strength of the perturbation $\lambda$.
In fact, for large dimensions the EA SF in the EGOE$(1+2)$ shows also a transition from a Breit-Wigner (BW) form,
\begin{eqnarray}
\label{breitwigner}
\mbox{SF}_{\rm EA}(w,E) &\approx& \mbox{SF}^{\rm BW}(w,E) \nonumber \\
&:=& \frac{1}{2\pi} \frac{{\Gamma}}{[(w+E-\bar \varepsilon)^2 + \frac{\Gamma^2}{4}]} \;,
\end{eqnarray}
for $\lambda < \lambda_F$, to a Gaussian form,
\bea
\label{SFgaussian}
\mbox{SF}_{\rm EA}(w,E)\approx \mbox{SF}^{\rm G}(w,E) := \frac{1}{\sqrt{2\pi\bar{\sigma}^2}}
e^{-[(w+E-\bar \varepsilon)^2/2\bar{\sigma}^2]} \;, \nonumber\\
\eea
for $\lambda > \lambda_F$ \cite{Kota2001report,Kotabook}, where $\bar \varepsilon := \bar \varepsilon(E)$, 
$\Gamma:=\Gamma(E)$, and $\bar \sigma:=\bar \sigma(E)$. We remark that the EGOE$(1 + 2)$ displays no degeneracies, so for such
ensemble the sums over $\alpha$ and $\gamma$ can be neglected (for instance, in Eqs. \eqref{PEAw} and \eqref{SFEA}).
\par
In the EGOE$(1+2)$ with large values of $N$, the ergodic property of both the level density and the strength function
have been verified \cite{Kota2001report,Kotabook}. Therefore, the  running average of such quantities is close to 
the ensemble average. For the strength function, it corresponds to 
$\mbox{SF}_{n}(w)\approx \mbox{SF}_{\rm EA}(w,E_n) $ in the limit $N\rightarrow \infty$. 
Usually, the approximation is verified for all perturbation strengths by performing
a histogram function of $\mbox{SF}_{n}(w)$ within a small energy window  
$E_{n_l}\leq E_n \leq  E_{n_r}$ (with $n_l \leq n \leq n_r$) and with a small binning of the variable $w$. By increasing 
the dimension of the matrices and maintaining the bin's size of the histogram, it is possible to verify  
the convergence  of the distribution $\mbox{SF}_{n}(w)$ to $\mbox{SF}_{\rm EA}(w,E_n)$ for every value of $E\sim E_n\,$. 
\par
In addition to the value of $\lambda_F$, the transition from an integrable to a chaotic regime in the EGOE(1+2) 
is determined by another value
of the interaction strength, $\lambda_c$, that sets the transition from a Poissonian to a Wigner-Dyson distribution 
in the nearest neighbor level spacing \cite{Kotabook}. Thus, these types of random ensembles are well suited to describe 
deterministic many-body interacting systems that present this type of transition in the level statistics~\cite{Santos2012,Borgonovi2016}. 

\section{Work distribution in many-body systems and random ensembles} 
\label{SectionV}

In Sec. \ref{section-random-models} we have shown that certain random models can be used to 
approximate the work statistics of complex systems. This is done by considering a smoothed version of the level density and of the
strength function, both obtainable through an ensemble average.
However, it is not yet clear the role played by the temperature and by the perturbation strength for an accurate description 
of the exact $P(w)$.
In the next sections we consider sudden quenches for two different one-dimensional spin-$1/2$ chains 
with Hamiltonians of the form of Eq.~\eqref{H+V}. In both systems, we compare the  
exact work pdf with the one given by the EGOE$(1+2)$ using the level density of Eq.\eqref{dens1} and the SF 
of Eqs.~\eqref{breitwigner} and \eqref{SFgaussian}
for a broad range of temperatures and perturbation strengths.
\begin{widetext}
\begin{figure*}[h!]
\centering 
\includegraphics[width=16cm]{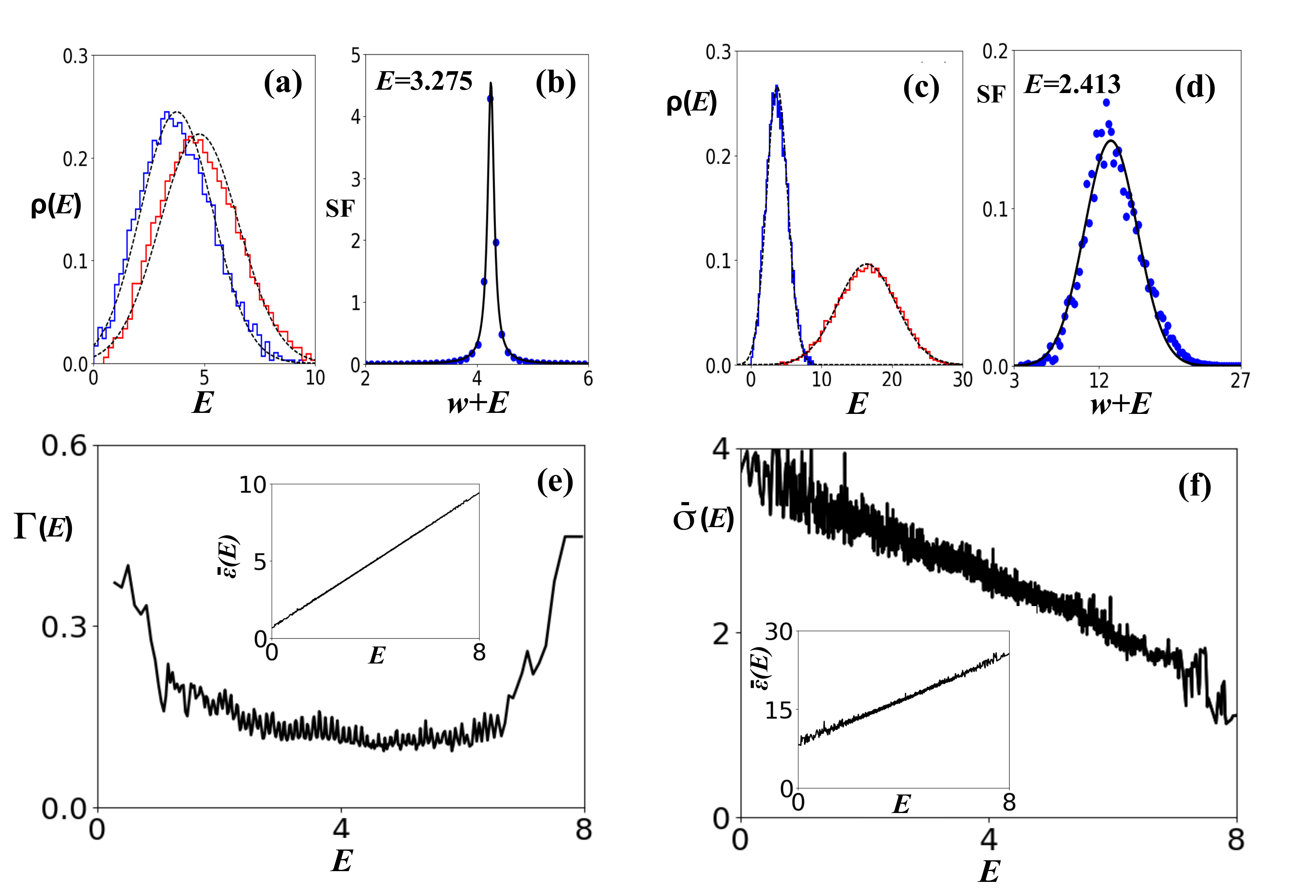}
\caption{(color online) 
Elements characterizing the work pdf of the model of Eq. \eqref{model2} 
with $\mu = 0.5$. Panels (a), (b), and (e) for a quench $\lambda_i=0.7\rightarrow \lambda_f=0.9 $; and (c), (d), and (f) 
for $\lambda_i=0.7\rightarrow \lambda_f=3.2$.
Panels (a) and (c): Histograms of the level densities. In (a) the left histogram (blue) is for $\rho(E)$ with $\lambda = 0.7$, 
and the right one (red) is for $\tilde \rho(\tilde E = E)$ with $\lambda = 0.9$. In panel (c) the histograms 
correspond to $\rho(E)$ with $\lambda = 0.7$, and  
$\tilde \rho(\tilde E = E)$ with  $\lambda = 3.2$. The black dashed curves are the Gaussian fittings in  Eq. \eqref{dens1}. 
Panels (b) and (d): the strength functions. 
Dots represent the histograms and the full black lines correspond to the Breit-Wigner fit in (b), $\mbox{SF}^{\rm BW}(w, E)$,
and to the Gaussian fit, $\mbox{SF}^{\rm G}(w,E)$, in (d). 
Panels (e) and (f):  In (e) we plot the width $\Gamma$ and the centroid $\bar{\varepsilon}$ (inset) of the fitting
$\mbox{SF}^{\rm BW}(w, E)$ as a function of the initial energy.  These values were obtained through the 
fitting of the corresponding histograms of the energy-smoothed strength function $\mbox{SF}(w, E)$ for 
different values of $E$. In  (f) we plot  the centroid $\bar{\varepsilon}$
and the width $\bar{\sigma}$ using Eqs. \eqref{centespec} and \eqref{varespec}, respectively. See text for details.}
\label{fig12} 
\end{figure*}
\end{widetext}

\subsection{Spin-1/2 chains}
\label{section-models}
\par
The models we considered are two different one-dimensional spin-$1/2$ chains with two-body finite-range interactions. 
The first one has only first neighbor interactions, and is given by \cite{Santos2012}
\bea
H_1 &:=& H_0 + \mu V_1 \,,\nonumber \\
H_0 &:=& \sum\limits_{i=1}^{L-1} J(S_{i}^{x}S_{i+1}^{x} + S_{i}^{y}S_{i+1}^{y}) \,,    \nonumber \\
V_1 &:=& \sum\limits_{i=1}^{L-1} J S_{i}^{z}S_{i+1}^{z}\;,
\label{model1}
\eea
where $L$ is the number of sites, and $S_{i}^{k}$, with $k = x$, $y$, or $z$, are the spin operators at site $i$.
The parameter $J$ sets the energy scale and is chosen to be $1$ in what follows. 
The flip-flop term, $H_0$, that moves the excitations (spin up in the $z$ direction) through the chain, can be mapped 
onto a system of noninteracting spinless fermions or hard-core bosons and is integrable.
Perturbations over $H_1$, known as the $XXZ$ Hamiltonian,  are applied by changing the coupling or anisotropy parameter, 
$\mu$, from some initial value $\mu_i$ to some final value $\mu_f$. 
Here we consider quenches that are obtained by a sudden  change $\mu_i \rightarrow \mu_f$. 
It is important to remark that the Hamiltonian $H_1$ is always integrable, irrespective to the 
strength of the parameter $\mu$.
\par

\begin{figure}[h!]
\centering 
\includegraphics[scale=0.4]{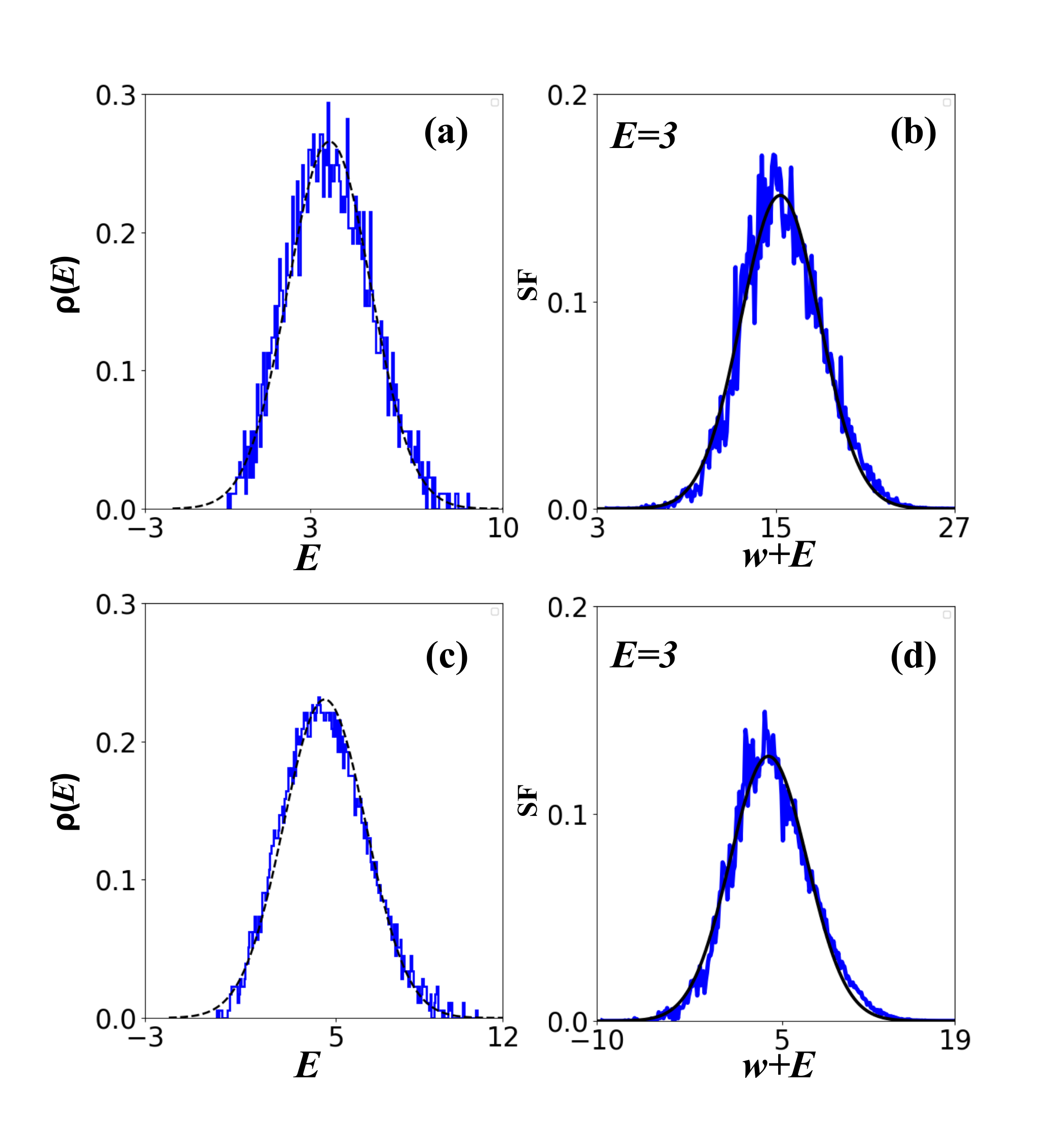}
\caption{(color online) For the same big quench of Fig. \eqref{fig12}, 
$\lambda_i=0.7\rightarrow \lambda_f=3.2$, 
we plot the histograms (in blue) of the initial level densities $\rho(E)$  ((a) and (c)), and the 
smoothed strength function $\mbox{SF}(w, E = 3)$ 
((b) and (d)). In panels (a)
and (b) the initial and final Hamiltonians have a dimension $N = 1512$ and 
in (c) and (d), $N = 3215\,$. As the dimension increases, the fluctuations of the histograms decrease 
around the Gaussian fittings. We kept the bin size fixed: in  (a) and (c) it is $0.055$ and in  
(b) and (d) it is $0.097\,$. See text for details.}
\label{fig3} 
\end{figure}
\par
The second model we consider is built by introducing second neighbor interactions to the first model, 
and is given by \cite{Santos2012} 
\bea
H_2 &:=& H_1 + \lambda V_2 \,, \nonumber \\
V_2 &:=& \sum\limits_{i=1}^{L-2} J[(S_{i}^{x}S_{i+2}^{x} + S_{i}^{y}S_{i+2}^{y}) + \mu S_{i}^{z}S_{i+2}^{z}] \;.
\label{model2}
\eea
The  quench we study in this system is given by a sudden change of the strength in the second neighbor 
interactions, $\lambda_i \rightarrow \lambda_f$.
The inclusion of second neighbor interactions,
given by $V_2$, allows the system to
display a chaotic structure when the value of $\lambda$ is sufficiently high \cite{Santos2012}. 
The momentum conservation is avoided by considering open boundary conditions.
However, other symmetries are present.
First, the two models  
conserve the total spin in the $z$ direction for any value of the parameters, viz., $[H_{1,2}, S^{z}] = 0$, with 
$S^{z} = \sum_{i = 1}^{L} S_{i}^{z}$. This conserved quantity allows us to break up
the total state space into subspaces of a fixed number $K$ of spins up which do not mix under evolution, and 
we work within such subspaces of dimension $D_K=L!/(K!\,(L-K)!)$.
The models also preserve the value of the total spin, $S^2=(\sum_{i=1}^L \vec{S}_{i})^2\,$, 
if $\mu=1$, then we do not use this value throughout. 
There is also a parity symmetry defined as the collective permutation of mirrored sites in the chain 
that is avoided by dealing with one of the two (positive or negative) parity subspaces. 
Therefore, the effective dimension of the systems considered is $N\approx D_K/2$.  
A systematic study of the level densities and the strength functions of the models 
in Eqs. \eqref{model1} and \eqref{model2} were performed in Ref. \cite{Santos2012}.
In order to make our presentation self-contained, we summarize the results of Ref. \cite{Santos2012}, and apply them
to sudden quench processes discussed here. 
It is worth mentioning that our results are equivalent to those of Ref. \cite{Santos2012}
if we consider, in each quench process $\lambda_i \rightarrow \lambda _f$ 
or $\mu_i \rightarrow \mu_f $, the eigenstates of the Hamiltonians 
$H_1 + \lambda_i V_2$ or $H_0 + \mu_i V_1$ as the mean-field basis, respectively.
On the contrary, in Ref. \cite{Santos2012} the mean-field bases were always the eigenstates 
of $H_1$ or $H_0\,$.
\par
Let us start by considering the level densities in Fig.~\ref{fig12} (panels (a) and (c)) 
that refer to Hamiltonians of the model in Eq. \eqref{model2}
with $L = 15$, $K = 5$, and $\mu = 0.5$.  Thus, in this case the dimension of the system we consider  is $N= 1512$.
In this system, there is a crossover from integrability to chaos that occurs for $\lambda_{c} \approx 0.5$, where the 
level spacing distribution changes from a Poissonian to a Wigner-Dyson distribution
\cite{Santos2012}.
The parameters considered in Fig. \ref{fig12} panel (a) are associated to small a quench ${\lambda_i = 0.7 \rightarrow \lambda_f = 0.9\,}$,
and the ones in panel (d)  to
a larger quench, ${\lambda_i = 0.7 \rightarrow \lambda_f = 3.2}$.
By small or large quench, we mean the value of the difference between the initial and final values of the coupling parameters
($\mu_f - \mu_i$ for the first model and $\lambda_f - \lambda_i$ for the second model).  If the shape of the SF is a Breit-Wigner distribution,
then we call it a small quench, and we call it a large quench if the SF is a Gaussian distribution.
Therefore, the  initial and final Hamiltonians  displayed in the figures are all in the chaotic regime.
The initial and final level densities, $\rho(E)$ and $\tilde \rho(\tilde E)$,
are shown in the form of histograms with small binning size.
The dashed curves show that, in both perturbation regimes, the histograms are well approximated by 
Gaussian distributions, $\rho_{\rm E}^{\rm G}(E)$, 
corresponding to the EA of the EGOE$(1+2)$, Eq. \eqref{dens1}.
\par
For the quench $\lambda_i = 0.7 \rightarrow \lambda_f = 0.9$ a typical strength function is  displayed in Fig. \ref{fig12}, panel (b).
In this case the SF is expected to behave as a Breit-Wigner distribution in the limit of large dimension $N$. 
This is verified in panel (b) of Fig.~\ref{fig12}, 
where we plot the energy-smoothed strength function $\mbox{SF}(w,E)$ as a 
histogram constructed from $\mbox{SF}_n(w)$ in an small
energy window  $E_{n_l}  \leq E_n \leq E_{n_r}$ around a fixed value 
of the initial energy $E\sim E_n$, and for a small bin size in the variable $w$. 
We fit this histogram with the EA  function, $\mbox{SF}^{\rm BW}(w,E)$ in Eq. \eqref{breitwigner}, of the EGOE$(1+2)$.
We have similar fittings when repeating the procedure but varying the initial energy $E\sim E_n$, 
and in panel (e) of Fig.~\ref{fig12} 
we show the center $\bar{\varepsilon}$ (inset) and the width of these Breit-Wigner fittings, viz. 
$\mbox{SF}^{\rm BW}(w,E)$ in Eq. \eqref{breitwigner},
as a function of the initial values of the energy $E$. 
On the other hand, for a  quench $\lambda_i = 0.7 \rightarrow \lambda_f = 3.2$, the SFs
behave as Gaussian distributions in the limit of large dimension $N$, then this perturbation is considered large. 
This is verified in panel (d) of Fig.~\ref{fig12}, where a Gaussian fitting is displayed. We remark that in this case
the centroid $\bar{\varepsilon}$
and the width $\bar{\sigma}$ of the SFs were computed from the matrix elements 
(without any diagonalization or fitting) using Eqs. \eqref{centespec} and \eqref{varespec},
and showed perfect agreement with those obtained by numerical diagonalization.
In panel (f) of Fig.~\ref{fig12} we show the center $\bar{\varepsilon}$ and the width of the $\mbox{SF}^{\rm G}(w, E)$, 
in Eq. \eqref{SFgaussian},
as a function of the initial values of the energy $E$. 
\begin{figure}[h!]
\centering 
\includegraphics[scale=0.4]{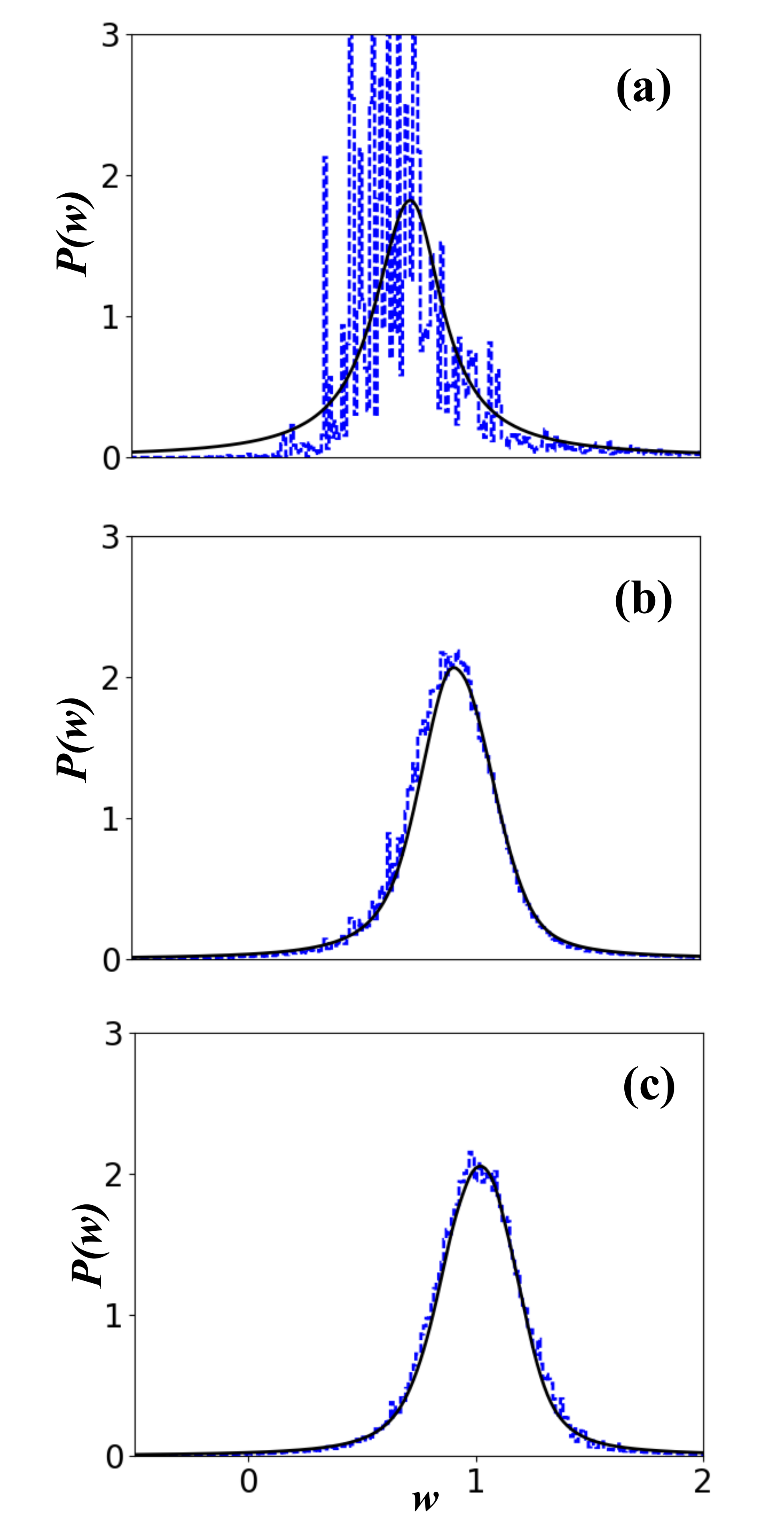}
\caption{(color online) The dashed lines (in blue) are the histograms that represent $P(w)$ in Eq. \eqref{PWquench}
for a small quench, $\lambda_i = 0.7 \rightarrow \lambda_f = 0.9$ , using the model in Eq. \eqref{model2} (with $\mu=0.5$ and $N=1512$) 
for inverse temperature values: (a) $\beta = 5$, so $N_{\rm eff}/N = 0.066\,$; (b) $\beta = 0.5$, so $N_{\rm eff}/N = 0.66$;
and (c) $\beta = 0.05$, so $N_{\rm eff}/N = 6.6\,$. The initial and final Hamiltonians are in the chaotic regime.
The bin size is $\sim 0.08$ and the total number of bins is
$1800$ in all the histograms. The full line (in black) corresponds to the EA, $P_{\rm EA}(w)$, in Eq. \eqref{PEAw}
using the Gaussian level density 
$\rho_{\rm E}^{\rm G}(E)$ in Eq. \eqref{dens1} and 
the Breit-Wigner function
$ \mbox{SF}^{\rm BW}(w,E)$ in Eq. \eqref{breitwigner} (see text for details). 
}
\label{fig4} 
\end{figure}

\begin{figure}[h!]
\centering 
\includegraphics[scale=0.4]{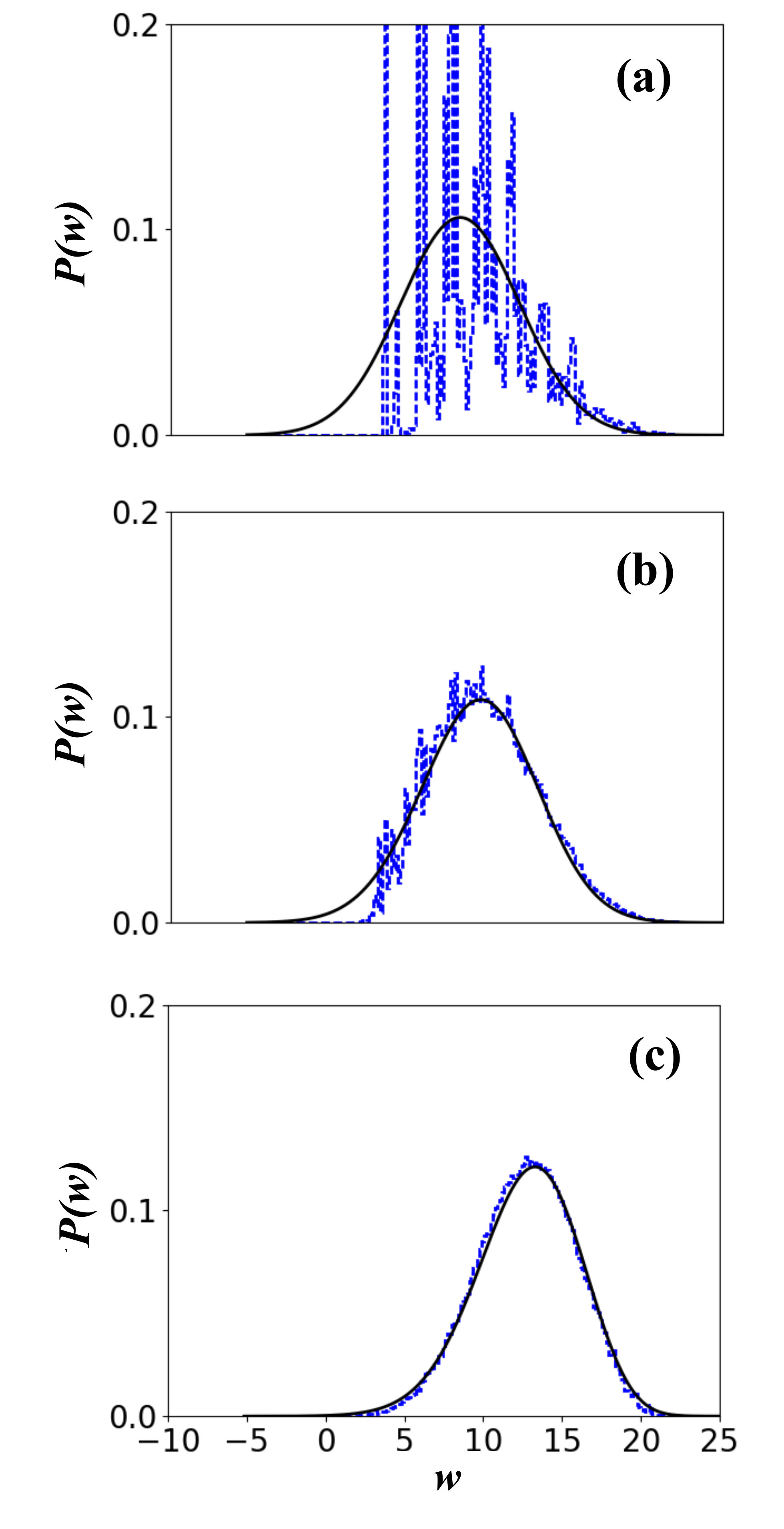}
\caption{(color online) 
The same as Fig. \eqref{fig4} but 
for a large quench, $\lambda_i = 0.7 \rightarrow \lambda_f = 3.2$ and for different values of the inverse of temperature: 
(a) $\beta = 20$, so $N_{\rm eff}/N = 0.0165$; (b) $\beta=2$, so $N_{\rm eff}/N = 0.165\,$;
and (c) $\beta = 0.005$, so $N_{\rm eff}/N = 66$. Here, both the initial and final Hamiltonians are in the chaotic regime.
The bin size is $\sim 0.14$ and the total number of bins is
$250$ in all histograms. The full line (in black) corresponds to the EA , $P_{\rm EA}(w)$ in Eq. \eqref{PEAw},
constructed using the Gaussian functions, $\rho_{\rm E}^G(E)$ in Eq. \eqref{dens1} and
$ \mbox{SF}^{\rm G}(w,E)$ in Eq. \eqref{SFgaussian} (see text for details).}
\label{fig5} 
\end{figure}
\par
We also consider quenches  in the model of Eq. \eqref{model1}, where the system is integrable. 
The first one corresponds to the change $\mu_i = 0.1 \rightarrow \mu_f = 0.5$ (that we call ``small perturbation''),
and the second to the change $\mu_i = 0.1 \rightarrow \mu_f = 2.4$ (that we call  ``large perturbation'').
In both cases the results (not shown) are analogous to those shown in Fig. \ref{fig12}, i.e.,
the initial and final level densities are well fitted by Gaussian distributions, $\rho_E^{\rm G}(E)$ in Eq. \eqref{dens1}, 
corresponding to the EA of an EGOE$(1+2)$. 
Unlike the EGOE$(1+2)$, the model of Eq. \eqref{model1}
does not have a transition to a chaotic regime. However,  we observe that the strength functions also behave 
as the prediction for these ensembles, i.e., for a small perturbation it follows the Breit-Wigner 
fitting, $\mbox{SF}^{\rm BW}(w, E)$ in Eq. \eqref{breitwigner}, and for a large perturbation, it follows the Gaussian 
fitting, $\mbox{SF}^{\rm G}(w, E)$ in Eq. \eqref{SFgaussian}. 
\par
We have checked the convergence of the level densities and strength functions
and found that ${\rho(E) \approx \rho^{\rm G}(E)}$ in Eq.~\eqref{dens1} for all ranges of parameters analyzed.
Similarly,  ${\mbox{SF}(w,E) \approx \mbox{SF}^{\rm BW}(w,E)}$
when small perturbations are considered, and ${\mbox{SF}(w,E) \approx \mbox{SF}^{\rm G}(w,E)}$
for large perturbations, independently of whether the initial and final Hamiltonians are in the chaotic or in the integrable regimes.
Convergence to the EA is shown in the histograms of Fig.~\ref{fig3} for the same sudden quench used in Fig. \ref{fig12} 
(panels (d), (f) and (g) ), but for two different  
dimensions: $N = 1512$ in panels (a) and (b) and $N = 3215$ in panels (c) and (d).
In these plots, one can see that by fixing the size of the histogram binning, the size of the fluctuations 
of $\rho(E)$ and $\mbox{SF}(w,E)$ around the curves $\rho^{\rm G}(E)$ and $ \mbox{SF}^{\rm G}(w,E)$, respectively, 
decreases as the dimension increases. 
\par
\begin{figure}[h!]
\centering 
\includegraphics[scale=0.23]{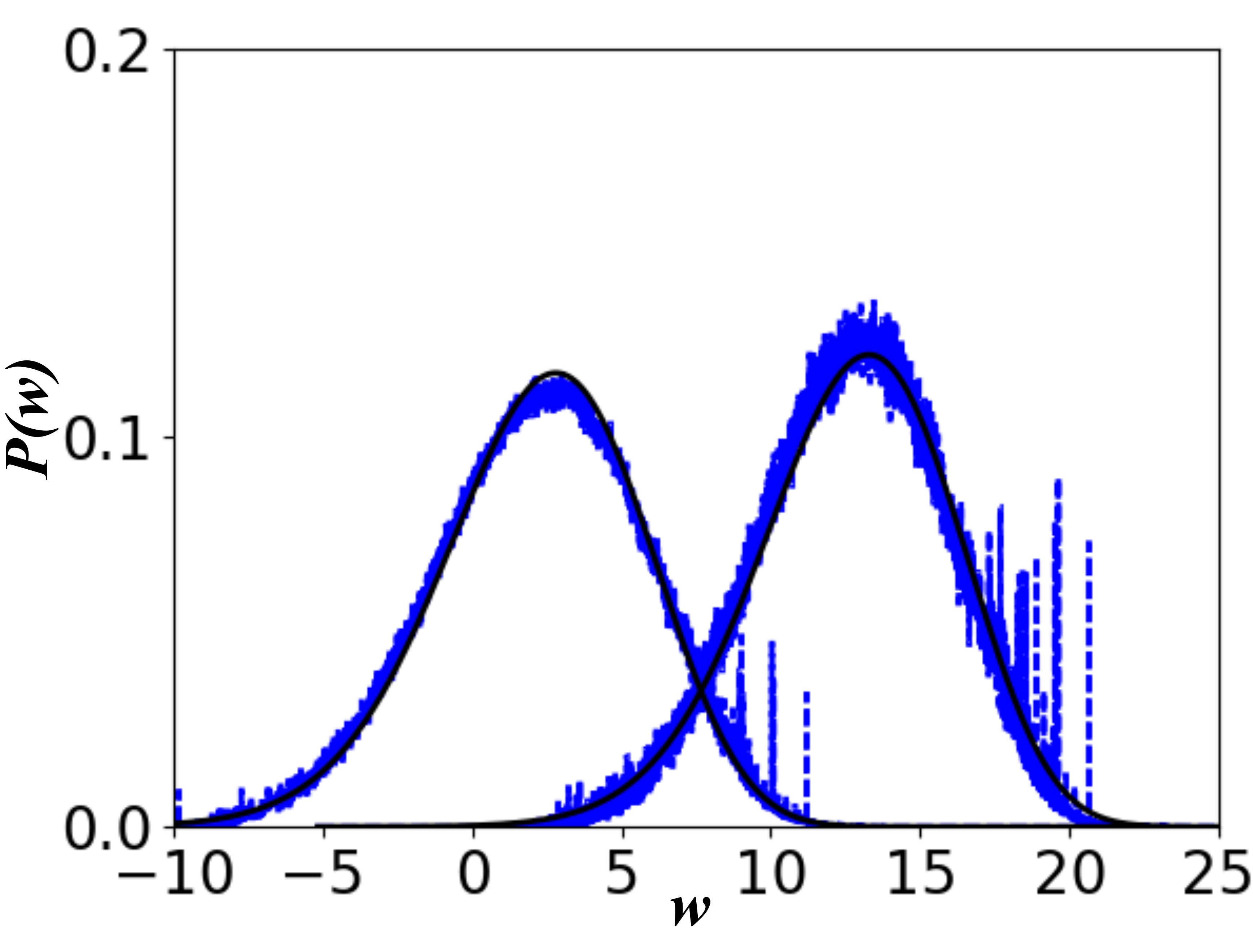}
\caption{(color online) The dashed lines (in blue) are the histograms that represent the exact work pdf $P(w)$ in 
Eq. \eqref{PWquench}
for a small quench, $\lambda_i = 0.7\rightarrow \lambda_f = 0.9$ , using the model in Eq. \eqref{model2} (with $\mu = 0.5$). 
We use a system with  $N = 1512$ for the plots on the right and 
$N = 3235$  for the plots on the left. In both cases $N_{\rm eff}/N = 60$.  The full lines (in black) correspond 
to the EA, $P_{\rm EA}(w)$, 
in Eq. \eqref{PEAw}
using the Gaussian level density 
$\rho_{\rm E}^{\rm G}(E)$ in Eq. \eqref{dens1} and 
$ \mbox{SF}^{\rm BW}(w,E)$ in Eq. \eqref{breitwigner}.}
\label{fig6} 
\end{figure}
\par
\begin{figure}[h!]
\centering 
\includegraphics[scale=0.4]{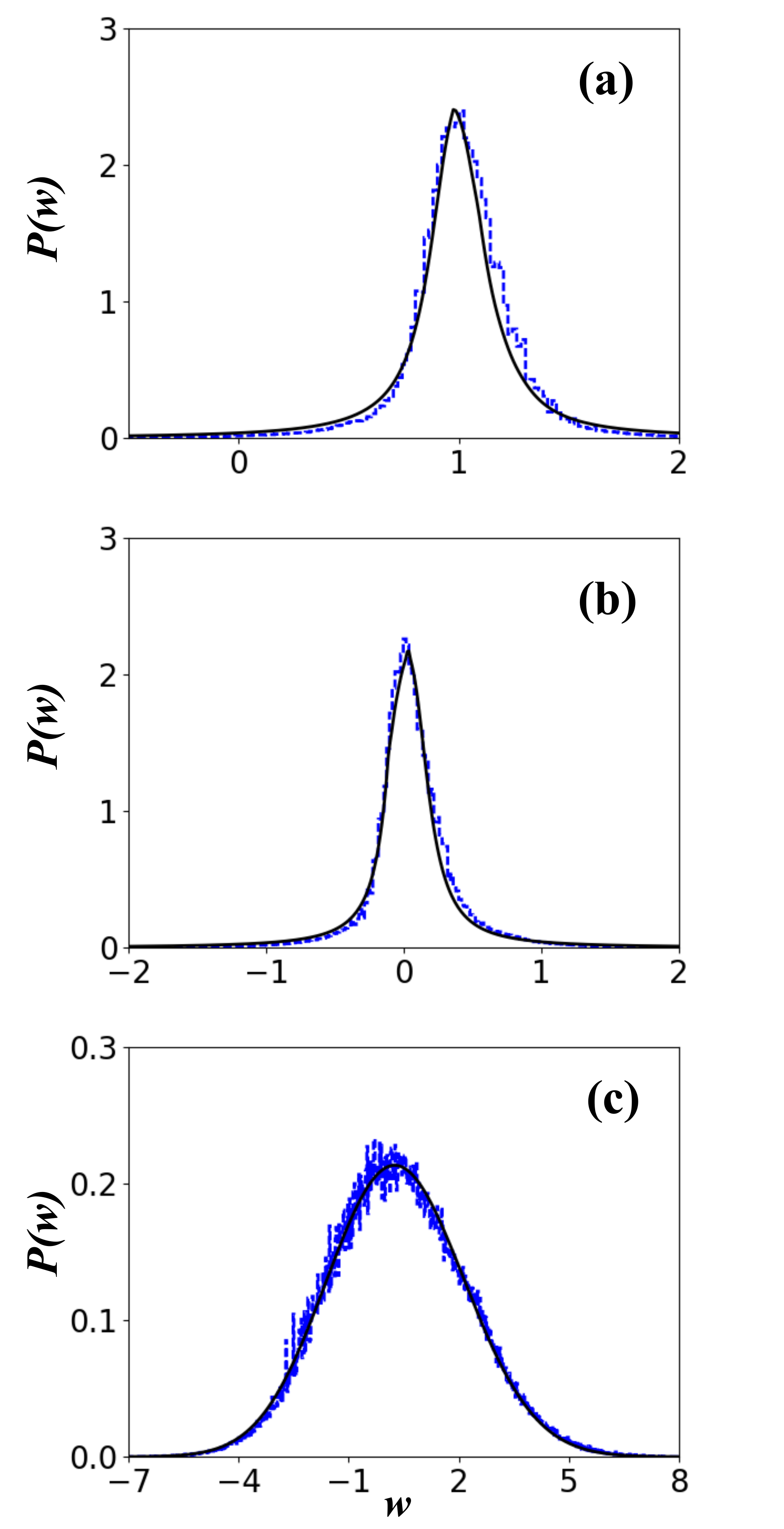}
\caption{(color online) The dashed lines (in blue) are the histogram functions that represent the exact work pdf, $P(w)$, 
in Eq. \eqref{PWquench} in the regime of high temperature ($\beta = 0.05$) and when the initial and final Hamiltonians  
are in the integrable regime. 
The dimension of Hamiltonian matrices is $N = 1512$ in all the systems used.
Therefore, $N_{\rm eff}/N = 8$ in all the panels.
In panel (a) we use the model in Eq. \eqref{model2} with a small quench, $\lambda_i = 0.1 \rightarrow \lambda_f = 0.3$. 
In panels (b) and (c) we use the model in Eq. \eqref{model1}. In (b) we have a small quench $\mu_i = 0.1 \rightarrow \mu_f = 0.5\,$. 
In (c) we have a large quench $\mu_i = 0.1 \rightarrow \mu_f = 2.4\,$.
All histograms have a bin size $\sim 0.02$, and a total number of bins $800$. The full lines (in black) correspond to the EA, $P_{\rm EA}(w)$, 
in Eq. \eqref{PEAw}
using the Gaussian level density 
$\rho_{\rm E}^{\rm G}(E)$ in Eq. \eqref{dens1} and 
the Breit-Wigner function
$ \mbox{SF}^{\rm BW}(w,E)$ in Eq. \eqref{breitwigner}  for panels (a) and (b) and $ \mbox{SF}^{\rm G}(w,E)$ in Eq. \eqref{SFgaussian} for panel (c).
} 
\label{fig7} 
\end{figure}
\par
\par
\subsection{Work distribution for spin chains}
\label{section-wpdf-models}

In the previous section we showed that the $\rho(E)$ and $\mbox{SF}(w,E)$ of the models considered can be borrowed from the EGOE$(1+2)$
when the dimension of the system is large, viz., $N \gg 1$. Both functions are fundamental pieces in the construction of
$P_{\rm sm}(w)$. However, the work pdf depends on the temperature through the Boltzman factor, and the accuracy
of $P_{\rm sm}(w)$ in describing the exact $P(w)$ must be checked for different ranges of temperatures.
Here, we show that the exact $P(w)$ can be described by
$P_{\rm sm}(w)$, or essentially by $P_{\rm EA}(w)$ (Eq.\eqref{PEAw} from EGOE$(1+2)$ models), only for
intermediate and large temperatures, being also analytically computable.
\par
The presence of the Boltzman factor in Eq. \eqref{PEAw}, or equivalently in Eq. \eqref{PWfinal},
allows the characterization of low or high temperatures by defining the parameter 
$N_{\rm eff}$ that is proportional to the number 
of levels below the energy $E$ such that $\beta E\sim 1$. Therefore, a given value of $\beta$ corresponds 
to a high temperature if $N_{\rm eff}/N \gg 1$ and to a low temperature if
$N_{\rm eff}/N \ll 1$.  We can roughly estimate $N_{\rm eff}$ by writing  
$\beta E \approx \beta N_{\rm eff} \,\bar s \approx 1$, with a mean level spacing 
approximately given by $\bar s=1/\rho_{\rm E}^{\rm G}(\bar E)$, with $\rho_{\rm E}^{\rm G}(\bar E)$ being the maximum value of the level 
density in Eq. \eqref{dens1}.
\par
In Figs. \eqref{fig4} and \eqref{fig5} we show the results for the same parameters that we considered in Fig. \eqref{fig12}.
The initial and final Hamiltonians are both in the chaotic regime. 
For intermediate and high temperatures, panels (b) and (c), respectively, we see a very good agreement between 
$P_{\rm sm}(w)\approx P_{\rm EA}(w)$ and $P(w)$, the latter represented by a histogram. 
However, for small temperatures, panel (a), we observe large deviations.
The reason is that for small temperatures $P(w)$ does not converge to a smooth function.
We check this by increasing the system's dimension $N$, but maintaining $N_{\rm eff}/N \ll 1$ fixed, 
and performing a histogram representing $P(w)$ with the same bin size as in Figs. \eqref{fig4} and \eqref{fig5}. 
On the one hand, we observe that the size of the 
fluctuations of the histograms does not decrease (plot not shown) in the case of low temperatures.  
On the other hand, the fluctuations of the histograms representing $P(w)$
decrease around the smooth curve for intermediate and high temperatures.
An example of this type of calculation is shown in Fig. \eqref{fig6}.
\par
In Fig. \eqref{fig7} we show that there is also a good agreement in the intermediate and high temperature regimes when the initial and final 
Hamiltonians are integrable, for both large and small perturbations.
\par
\section{Conclusions}
\label{SectionVI}
\par
We have shown that for quench processes in quantum systems with large Hilbert spaces, such as interacting quantum many-body systems, 
there is a simple way to describe the work distribution function.
Such construction is a smoothed work pdf given by an energy integration of the 
product of two energy smooth functions, weighted by a Boltzmann factor. The smooth
functions are
the level density 
of the initial Hamiltonian and the strength function of the
eigenstates of the initial Hamiltonian in the eigenbasis of the final one. 
\par
We also have shown that an equivalent expression
can be obtained for the ensemble average of the work distribution for quantum quenches over random Hamiltonian models, provided that the 
eigenvalues and eigenvectors of the Hamiltonians in the ensembles are statistically independent, and 
also that an annealing approximation over the ensemble average is valid. 
The latter condition is usually verified when the dimension of the ensemble matrices are sufficiently large. 
When the ensembles of random matrices describe well the density of levels and the 
corresponding strength function of the systems considered, 
the energy-smoothed work pdf can be obtained 
from the ensemble average.
This is very advantageous since, in general, this approach provides 
analytical expressions valid for many realistic systems.
\par 

We numerically checked in spin-$1/2$ chains models, whose level 
density and strength function are well described by the EGOE$(1+2)$, that the  
exact quantum work pdf has small fluctuations around the energy-smoothed work pdf for large
Hilbert space dimensions.
Our results show that the energy-smoothed work pdf represents a good description of the exact work pdf for 
intermediate and high temperatures, but fails in the regime of low temperatures.
We have also verified the agreement between the smoothed description and the exact work pdf 
in both integrable and chaotic regimes for large and small quench strengths.

The approach developed here is general and 
avoids the demanding task of diagonalization of many-body interacting Hamiltonians. 
Our findings constitute one step further towards a generic description of the thermodynamics of many-body Hamiltonians, with
possible implications in quantum chaos and condensed matter physics.


\acknowledgments{E.G.A. and F.T. acknowledge financial support from the Brazilian funding agencies CNPq, 
CAPES (PROCAD2013 project), FAPERJ, and the National Institute of Science and Technology - Quantum Information.
D.A.W. and A.J.R. have received funding from CONICET (Grant No. PIP 11220150100493CO) ANPCyT(Grants No. PICT- 2016-1056 and No. PICT 2014-3711), 
and UBACyT (Grant No 20020170100406BA).}\\

\appendix

\section{Ensemble average of work distribution in random models}
\label{appendixA}

This appendix shows the demonstration of Eq.(\ref{PEAw}).
After performing the EA 
of the expression in Eq. \eqref{PWquench}, where we used the linearity of averages, we see that the result 
does not depend on the indexes``$n$'' and ``$m$'', due to the first assumption of 
statistical equivalence of the eigenvalues and 
eigenvectors of the Hamiltonians. Therefore, we can write
\begin{widetext}
\bea
P_{\rm EA}(w)&=&N^2\left\langle\left\langle\left\langle\left\langle
\frac{e^{-\beta E}\left( \sum_{\alpha,\gamma}\left|\ip{\tilde\psi^{\alpha}(\tilde{\bf E},\tilde{\boldsymbol{\theta}})}{\psi^{\gamma}({\bf E},\boldsymbol{\theta})}\right|^2 \right)\delta(w-\tilde E+E)}{{\cal Z}_0({\bf E})}
\right\rangle\right\rangle_{{\bf E},\boldsymbol{\theta}}
\right\rangle \right\rangle_{\tilde{\bf E},\tilde{\boldsymbol{\theta}}}\nonumber\\
&\approx&N^2
\frac{\left\langle\left\langle\left\langle\left\langle
e^{-\beta E}\left( \sum_{\alpha,\gamma}\left|\ip{\tilde\psi^{\alpha}(\tilde{\bf E},\tilde{\boldsymbol{\theta}})}{\psi^{\gamma}({\bf E},\boldsymbol{\theta})}\right|^2 \right)\delta(w-\tilde E+E)
\right\rangle\right\rangle_{{\bf E},\boldsymbol{\theta}}
\right\rangle \right\rangle_{\tilde{\bf E},\tilde{\boldsymbol{\theta}}}}
{\left\langle
{\cal Z}_0({\bf E})
\right\rangle_{{\bf E}}} \,,
\label{annelingapprox}
\eea
\end{widetext}
where we have used  in the last line an annealing approximation \cite{Cotler2017, Arrais2018, Xu2018, Chenu2019}, constituting 
the second assumption. Such approximation
is valid, in principle, for high-dimensional random Hamiltonian models and
for any value of the inverse of temperature, $\beta$. In the denominator of 
Eq. \eqref{annelingapprox}, we have used also that 
$\langle\langle\expval{\expval{{\cal Z}_0({\bf E})}}_{{\bf E},\boldsymbol{\theta}}\rangle\rangle_{\tilde{\bf E},\tilde{\boldsymbol{\theta}}}=\expval{{\cal Z}_0({\bf E})}_{\bf E}$.
Further assuming that the joint distributions approximately factorize as independent functions of the eigenlevels 
and eigenvectors, viz.,
$P({\bf E}, {\boldsymbol{\theta}})\approx P({\bf E}) P( {\boldsymbol{\theta}})$ and 
$\tilde P(\tilde{\bf E},\tilde {\boldsymbol{\theta}})\approx \tilde P(\tilde{\bf E})\tilde P(\tilde {\boldsymbol{\theta}})$, 
it is easy to rearrange the expression in 
Eq.\eqref{annelingapprox} to obtain the expression in Eq.\eqref{PEAw}.

\section{Probability of work in Gaussian ensembles}
\label{Gaussian-ensembles}
\par
In this appendix we obtain
the results of \cite{Arrais2018} for Gaussian ensembles, but using Eq. \eqref{PEAw}.
For $N\gg 1$, the density of levels of
Hamiltonians from the Gaussian ensembles
follows the semicircle law \cite{Mehta-book, Weidenmuller2009}:
\beq
\label{sc-law}
\rho_{\rm E}(E)\approx \rho_{N\gg 1}(x)= \left\{ \begin{matrix} 
      \frac{2N}{\pi a}\sqrt{1-\left(\frac{x}{a}\right)^2}&, \mbox{
    $\frac{|x|}{a} \le 1$} \\
      0 &,  \mbox{
    $\frac{|x|}{a} > 1$} \\
   \end{matrix}\right.,
\eeq
with $x = E - \bar{E}$ and $a = 2N \bar{s}/\pi \,$.
The parameter $\bar{E}$ is the mean value of the Gaussian distributions of 
the independent random diagonal entries of the Hamiltonian matrices of the ensemble. Therefore,
it fixes the center of the
random matrix spectrum.
The semicircle behavior for large dimensions implies that the level spacing, $\bar{s}$, is almost constant for a large portion
of the spectrum, being almost equal to its value in the center:
\begin{equation}
\label{mspaceRMT}
\bar{s} = \frac{1}{ \rho_{N\gg 1} (0)}=
\pi \sigma\sqrt{\frac{\beta_e}{2N}}\,.
\end{equation} 
The parameter $\sigma^2$ is the variance of the diagonal as well as of the real and 
imaginary parts of the off-diagonal random elements Gaussian distributed \cite{Mehta-book}, setting the energy scale of the ensemble. 
The constant $\beta_e$ assumes the values $1$, $2$ for the GOE and GUE, respectively, 
and $\beta_e=4$ for the GSE.
\par 
For large dimensions we have:
\beq
\label{medvecRMT}
 \langle\langle
 |\ip*{\tilde{\psi}(w+E,\tilde{\boldsymbol{\theta})}}{\psi(E,\boldsymbol{\theta})}|^2
 \rangle_{\boldsymbol{\theta}}\rangle_{\tilde{\boldsymbol{\theta}}}\approx \frac{1}{a_{\beta_e}N} \,.
\eeq
The parameter $a_{\beta_e}$ is related to the degree of degeneracy of the ensembles, assuming the values
$a_{\beta_e} = 1$ for $\beta_e = 1$ and $2$, and $a_{\beta_e}=2$ for $\beta_e=4$. That is, while the GOE and GUE have
no degeneracies, the GSE spectrum is doubly degenerated.
Substituting Eqs. \eqref{sc-law} and \eqref{medvecRMT} in Eq. \eqref{SFEA}, we obtain
\begin{equation}\label{sfrmt}
\mbox{SF}_{\gamma,\rm EA}(w + E) \approx  \frac{\tilde{\rho}_{N \gg 1}(w + E)}{N} \;,
\end{equation}
where we recall that the sums over $\alpha$ and $\gamma$ run over the degeneracies (until $a_{\beta_e}$).
The EA of the partition function can be obtained from 
$\mathcal{Z}_{0, \rm EA}:=\int^\infty_{-\infty} dE g(E)\,e^{-\beta E}\,\rho_{\rm E}(E)\,$
by using Eq. \eqref{sc-law} and recalling that $g(E) = a_{\beta_e}$ for every $E$: 
\bea
\label{avpartition}
{\cal Z}_{0, \rm EA}
&\approx& a_{\beta_e} \int_{-\infty}^{\infty}d{E}\;e^{-\beta E}\, \rho_{N\gg 1}(E) \nonumber\\
&=& 2\;a_{\beta_e}\;e^{-\beta \langle E \rangle}\;\frac{I_{1}(2N \expval{ s} \beta/\pi)}{2\expval{ s} \beta/\pi},
\eea
with $I_n(x)$ being the modified Bessel function of first kind.
Thus, we finally obtain the EA of the work pdf:
\begin{eqnarray}
P_{\rm EA}^{\scriptsize{\rm GE}}(w) &=& \frac{a_{\beta_e}}{N\mathcal{Z}_{0, \rm EA}} \int_{-\infty}^{\infty} dE  
\rho_{N\gg 1}(E) \tilde{\rho}_{N \gg 1}(w + E)\;e^{-\beta E}\,, \nonumber \\
\label{PWfinalRMT}
\end{eqnarray}
where $ \rho_{N\gg 1}$ and $ \tilde{\rho}_{N\gg 1}$ are given by Eq. \eqref{sc-law} and ${\cal Z}_{0, \rm EA}$ is given by 
Eq. \eqref{avpartition}.
The expression for $P_{\rm EA}^{\scriptsize{\rm GE}}(w)$ in Eq. \eqref{PWfinalRMT} is recovered 
from the inverse Fourier transform of the EA of characteristic function, viz. $\expval{G(u)}$, given by Eq.(17) of \cite{Arrais2018}.
\par
Therefore, for Gaussian ensembles of large dimension, the EA of the work pdf for sudden quenches, $P_{\rm EA}^{\scriptsize{\rm GE}}(w)$, 
is simply given by the thermal average of the convolution of the level densities of the two random spectra. It is completely characterized  
by the average level spacings, $\bar{s}$ and $\bar{\tilde s}$, through Eq. \eqref{mspaceRMT}, and by the averages of the
eigenlevels, $\bar{E}$ and $\bar{\tilde E}$, corresponding to the center of the initial and final spectra, respectively.

\bibliographystyle{apsrev}
\bibliography{Master_Bibtex}

\end{document}